\nonstopmode

\newcommand{\ie}{i.e.{}}
\newcommand{\eg}{e.g.{}}
\newcommand{\eV}{\U{eV}}
\newcommand{\bohr}{\, a_0}

\newcommand{\U}[1]{\,{\rm{#1}}}
\newcommand{\X}[1]{_{\mathrm{#1}}}
\newcommand{\Int}{\int\limits}
\newcommand{\differential}{\>\mathrm d}
\newcommand{\E}[1]{\times 10^{#1}}
\newcommand{\XUV}{\textsc{xuv}}
\newcommand{\nbh}{\hbox{-}}
\newcommand{\xray}{x\nbh{}ray}
\newcommand{\Xray}{X\nbh{}ray}

\documentclass[aps,pra,10pt,twocolumn,showpacs,nopreprintnumbers,superscriptaddress,nolongbibliography,amsmath,amssymb]{revtex4-1}
\usepackage{graphicx}
\usepackage[nativepdf,bookmarks,bookmarksopen,bookmarksnumbered,raiselinks,breaklinks,%
pdftitle={Optical control of an atomic inner-shell x-ray laser},%
pdfauthor={Gábor Darvasi, Christoph H. Keitel, Christian Buth},%
pdfsubject={Atomic Physics, Optics},%
pdfkeywords={X rays, photoabsorption, Auger decay, light-matter interaction, neon, x-ray lasing, atomic inner shell x-ray laser, optical control, EIT for x rays, Autler-Townes doublet}]{hyperref}

\begin{document}
\title{Optical control of an atomic inner-shell x-ray laser}
\author{G\'abor Darvasi}
\affiliation{Max-Planck-Institut f\"ur Kernphysik, Saupfercheckweg~1,
69117~Heidelberg, Germany}
\author{Christoph H.~Keitel}
\affiliation{Max-Planck-Institut f\"ur Kernphysik, Saupfercheckweg~1,
69117~Heidelberg, Germany}
\author{Christian Buth}
\thanks{Corresponding author.
Present address: Theoretische Chemie, Physikalisch-Chemisches Institut,
Ruprecht-Karls-Universit\"at Heidelberg, Im Neuenheimer Feld~229,
69120~Heidelberg, Germany.
Electronic mail}
\email{christian.buth@web.de}
\affiliation{Max-Planck-Institut f\"ur Kernphysik, Saupfercheckweg~1,
69117~Heidelberg, Germany}
\affiliation{Argonne National Laboratory, Argonne, Illinois~60439, USA}
\date{18 October 2014}

\begin{abstract}
\Xray~free-electron lasers have had an enormous impact on \xray~science
by achieving femtosecond pulses with unprecedented intensities.
However, present-day facilities operating by the self-amplified spontaneous
emission~(SASE) principle have a number of shortcomings,
namely, their radiation has a chaotic pulse profile and short
coherence times.
We put forward a scheme for a neon-based atomic inner-shell
\xray~laser~(XRL) which produces temporally and spatially
coherent subfemtosecond pulses that are controlled by
and synchronized to an optical laser with femtosecond precision.
We envision that such an XRL will allow for numerous applications
such as nuclear quantum optics and the study of ultrafast
quantum dynamics of atoms, molecules, and condensed matter.
\end{abstract}

%
%
%

\preprint{arXiv:1303.2187}
\maketitle

\section{Introduction}

An \xray~laser~(XRL) is an old goal of laser
physics~\cite{Milonni:LP-10,Rocca:TT-99,*Suckewer:XR-09}.
Extending the superior coherence, intensity and controlled pulse
properties of lasers to the \xray~regime has the potential to
revolutionize \xray~science by bringing unprecedented intensities,
temporal and spatial control of beam properties, and ultrashort
pulses to the \xray~scientist.
Such an XRL would be suitable for numerous applications from ionic and
nuclear quantum optics~\cite{Adams:QO-13,*Rohlsberger:EI-12}
to high-energy physics and astrophysics~\cite{DiPiazza:EH-12,*Bernitt:LO-12}
and it offers perspectives for the measurement and control of quantum
dynamics of matter on an ultrafast time scale~\cite{Krausz:AP-09,%
*Cavaletto:FC-13,*Cavaletto:HF-14}.
\Xray~free-electron-lasers~(FELs)~\cite{Madey:SE-71,*Saldin:PF-00} such as the
Linac Coherent Light Source~(LCLS)~\cite{LCLS:CDR-02,*Emma:FL-10} have
reached several of the goals laid out for XRLs, namely, ultraintense,
tunable, femtosecond \xray~pulses.
However, present-day \xray~FELs, operating by the principle of
self-amplified spontaneous emission~(SASE)~\cite{Kondratenko:GC-79,%
*Bonifacio:CI-84,LCLS:CDR-02,Emma:FL-10}, suffer from a
number of shortcomings compared with optical lasers.
Specifically, they lack controllability of the pulse properties,
spectral narrowness, and photon energy stability~\cite{LCLS:CDR-02,Emma:FL-10}.
Furthermore, there is a jitter between FEL x~rays and an optical laser
which can only be determined with the precision of several tens of
femtoseconds~\cite{Bionta:SE-11,*Schorb:XR-12,*Harmand:FF-13}.
Yet there are exciting novel facilities such as
FERMI@Elettra~\cite{Allaria:HC-12}
and self-seeding at LCLS for hard x~rays~\cite{Amann:SE-12} becoming
available that use seeding to improve on the properties of the \xray~pulses
reducing their bandwidth and the fluctuations of the pulse shapes.

Lasing schemes from the optical regime cannot be transferred
simply to the \xray~regime due to the unfavorable scaling of the
cross section for stimulated emission~\cite{Milonni:LP-10},
the lack of high-reflectivity mirrors for
x~rays~\cite{Rocca:TT-99,Suckewer:XR-09},
and the short duration of population inversion caused by
inner-shell hole decay~\cite{Als-Nielsen:EM-01,Schmidt:ES-97}.
Lasing in the \XUV{} and soft \xray~regimes has been
accomplished with plasma-based XRLs via collisional or recombinational
pumping~\cite{Rocca:TT-99,Suckewer:XR-09}.
However, these techniques have not yet reached beyond the water
window starting at~$276.2 \eV$.

A scheme for \xray~lasing in the kiloelectronvolt regime was proposed by
Duguay and Rentzepis in the year~1967 based on \xray~emission from
inner-shell transitions in core-ionized atoms~\cite{Duguay:SA-67}
which may be produced by focusing an intense \xray~beam from a FEL
into a gas cell with
atoms~\cite{Lan:PP-04,Rohringer:AI-09,*Rohringer:ER-10,*Rohringer:LA-09}.
Here one exploits that, at a given photon energy above but close to an
inner-shell edge, these tightly bound inner-shell electrons are much more
likely to interact with FEL x~rays than electrons in other
shells~\cite{Als-Nielsen:EM-01}, and thus an inner-shell vacancy is produced
leaving the cation in a state of population inversion with respect to radiative
transitions of electrons from higher-lying shells into the vacancy.
\Xray~lasing may occur in a macroscopic medium of such cations, in analogy to
optical lasers, by the propagation of initially spontaneously emitted x~rays
through the medium leading to stimulated emission of x~rays.

In this work, we would like to present the optical control of a modified
Duguay-Rentzepis scheme for an atomic inner-shell XRL.
The \xray~lasing scheme is presented in Sec.~\ref{sec:scheme},
the small-signal gain~(SSG) and its dependence on the optical laser
intensity are discussed in Sec.~\ref{sec:SSG}, and
the propagation of the FEL x~rays, the optical laser,
and the XRL light are examined in Sec.~\ref{sec:propagation}.
Finally, conclusions are drawn in Sec.~\ref{sec:conclusion}.
Atomic units~\cite{Hartree:WM-28} are used throughout unless stated otherwise.

\section{X-ray lasing scheme}
\label{sec:scheme}

\begin{figure}[tb]
    \centering
    \includegraphics[width=\hsize,clip]{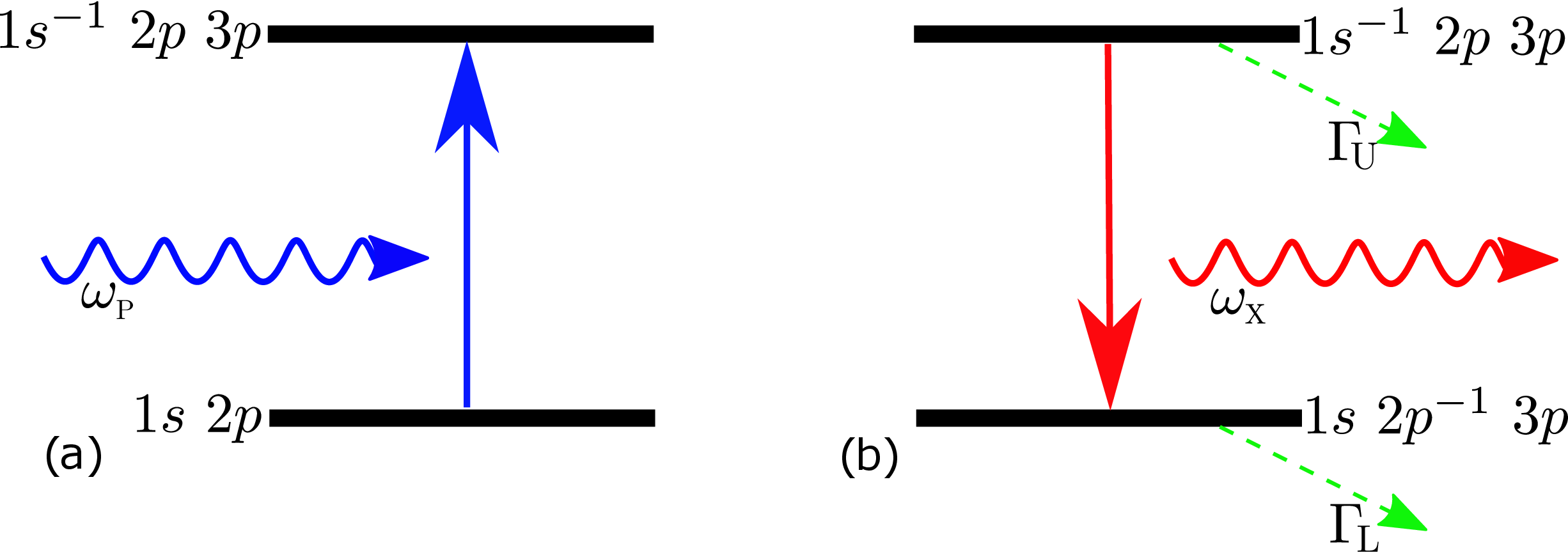}
    \caption{(Color online) Schematic of \xray~lasing in core-excited neon with
             pumping by an \xray~FEL.
             (a)~Neon atoms are excited by x~rays from a FEL from
             the ground state to the $1s^{-1} \, 2p \, 3p$~core-excited state.
             (b)~\Xray~lasing occurs on the $1s^{-1} \, 2p \, 3p \to
             1s \, 2p^{-1} \, 3p$~transition.
             Furthermore, $\Gamma\X{U}$ and $\Gamma\X{L}$
             are the decay widths of the upper and lower lasing
             states, respectively.}
    \label{fig:scheme}
\end{figure}

We develop a modified Duguay-Rentzepis scheme~\cite{Duguay:SA-67}
for which atoms are core excited instead of core ionized by FEL x~rays.
The principle is illustrated in Fig.~\ref{fig:scheme} by which \xray~lasing
proceeds as follows~\cite{Darvasi:OC-11}:
first, atoms are core excited by FEL \xray~absorption producing a state of
population inversion [Fig.~\ref{fig:scheme}a].
Second, photons are emitted through spontaneous radiative decay
of some of the core-excited atoms [Fig.~\ref{fig:scheme}b].
Third, a fraction of these photons copropagates with the FEL pulse
through the medium and induces stimulated \xray~emission from other
core-excited ions farther downstream in straight conceptual analogy to
optical lasers~\cite{Milonni:LP-10}.
Due to longitudinal pumping, \xray~lasing only occurs in the
propagation direction of the pump
pulse~\cite{Rohringer:AI-09,*Rohringer:ER-10,*Rohringer:LA-09,Darvasi:OC-11}.
Until recently it was not possible to realize such an XRL
due to the high pump \xray~intensity~\cite{Kapteyn:PI-89,*Kapteyn:PP-92}
that is necessary to excite the lasing medium faster than inner-shell holes
decay~\cite{Als-Nielsen:EM-01,Schmidt:ES-97}.

The advent of \XUV{} and \xray~FELs has reinvigorated interest in
Duguay-Rentzepis-style schemes, producing a number of theoretical
studies~\cite{Lan:PP-04,Rohringer:AI-09,Rohringer:ER-10,Rohringer:LA-09,%
Darvasi:OC-11} and the experimental realization for neon in the year~2011
at LCLS~\cite{Rohringer:AI-12}.
Although this simple, uncontrolled XRL~scheme produces pulses
with a single, fully coherent intensity spike, it still suffers
from a number of shortcomings: the pulse properties depend
on the temporal shape of the pump pulse and multicolor lasing
may occur~\cite{Rohringer:AI-09,Rohringer:ER-10,Rohringer:LA-09}
for suitably high FEL \xray~photon energies due to \xray~lasing on
transitions in highly charged ions.
Multiple uncontrolled pulses may lead to complications
in experiments harnessing the XRL~light.
However, multicolor \xray~lasing can be suppressed in the
Duguay-Rentzepis scheme with core ionization~\cite{Duguay:SA-67},
if the photon energy of the FEL x~rays is tuned only slightly above
the inner-shell absorption edge.
Then the FEL~photon energy is not high enough to core ionize cations;
yet core excitations of cations are energetically accessible, thus
leading to more than one XRL~transition~\cite{Oura:IS-10}.
Although multicolor \xray~lasing was predicted~\cite{Rohringer:AI-09,%
Rohringer:ER-10,Rohringer:LA-09}, it has not been observed experimentally
yet~\cite{Rohringer:AI-12}.

\begin{figure}
    \centering
    \includegraphics[width=\hsize,clip]{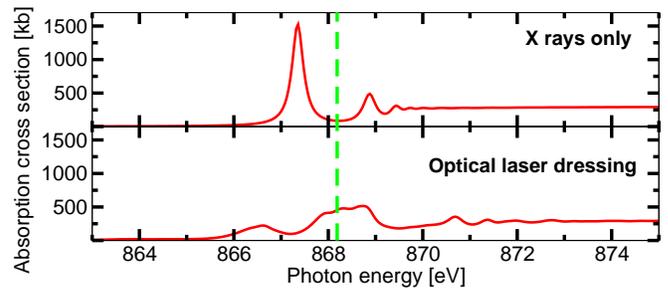}
    \caption{(Color online) \Xray~absorption cross section by core electrons
             of neon near the $K$~edge for the field-free case (top) and for
             optical laser dressing at~$10^{13} \U{W/cm^2}$ (bottom) with an
             $800 \U{nm}$~laser.
             The x~rays and the optical light are linearly polarized
             and have parallel polarization vectors.
             The dashed green line indicates the central FEL
             photon energy at~$\omega\X{P} = 868.2 \eV$.
             Adapted from Ref.~\onlinecite{Buth:ET-07}.}
    \label{fig:crosssection}
\end{figure}

We put forward an XRL~scheme for neon that is controlled by optical light.
The FEL \xray~photon energy, $\omega\X{P} = 868.2 \eV$, is tuned
below the $K$~edge slightly above the $1s \to 1s^{-1} \, 3p$~resonance
[Fig.~\ref{fig:scheme}], as indicated by the dashed green line in the
\xray~absorption cross section by core electrons of neon
in Fig.~\ref{fig:crosssection}.
We choose a FEL \xray~photon energy of~$\omega\X{P}$ also for the FEL
\xray-only case in order to facilitate comparison of the FEL \xray-only
case with the optically controlled XRL~case.
Namely, for this choice of~$\omega\X{P}$, the FEL \xray-only case represents the
optical-laser-off limit of the optically controlled XRL which
allows the strong increase of the small-signal gain with
the optical laser intensity as shown in Fig.~\ref{fig:combined} and
discussed in Sec.~\ref{sec:SSG}.
Hence, only in this case, strong optical control of the XRL is exerted.
For the FEL x~rays only, core excitation is off resonant and thus fairly
inefficient [Fig.~\ref{fig:crosssection}, top], if the bandwidth of
the FEL x~rays is limited sufficiently, \eg,
by seeding techniques~\cite{Amann:SE-12,Allaria:HC-12}.
\Xray~lasing occurs on the superposition of
$1s^{-1} \, 2p \, 3p \to 1s \, 2p^{-1} \, 3p$ and
$1s^{-1} \, 2p \, 4p \to 1s \, 2p^{-1} \, 4p$~transitions
leading to a two-peak feature around~$849 \eV$
for highly monochromatized x~rays~\cite{Oura:IS-10}.
%
%
We assume a full width at half maximum~(FWHM) bandwidth of the FEL x~rays
of~$\Delta \omega\X{P} = 0.4 \eV$ which is sufficiently broad such that
the two-peak structure disappears.

An optical laser with intensity~$I\X{L}$, that copropagates with the FEL
x~rays, modulates the \xray~absorption cross section by core electrons
of neon~$\sigma\X{X}(I\X{L}, \omega)$~\cite{Buth:ET-07,Santra:SF-07,%
Varma:XA-08,*Santra:SF-08,*Buth:RA-10,*Young:US-10,*Glover:CX-10}, as shown in
Fig.~\ref{fig:crosssection} (bottom), which thus becomes dependent on~$I\X{L}$.
The $\sigma\X{X}(I\X{L}, \omega\X{P})$~is increased by more than sixfold
from~86 to~$510 \U{kilobarn}$ when $I\X{L}$~is increased from~$0 \U{W/cm^2}$
to~$10^{13} \U{W/cm^2}$ for an optical dressing laser
with~$800 \U{nm}$~wavelength~\cite{Buth:ET-07}.
For this study, the core-excitation cross section~$\sigma\X{X}(I\X{L},
\omega)$ is calculated as a function of the \xray~photon energy~$\omega$
at nine different optical laser intensities~$I\X{L}$ with the
\textsc{dreyd}~program~\cite{Buth:AR-08,*fella:pgm-V1.3.0,%
Buth:TX-07,Buth:ET-07,Santra:SF-07,Buth:RA-10,Young:US-10,Gaarde:LD-11},
whereby the wavelength of the optical laser is not varied~%
\footnote{The optical-laser-intensity-dependent \xray~absorption cross
section of core electrons of neon is plotted in Fig.~2 of
Ref.~\onlinecite{Santra:SF-07} on the $1s \to 1s^{-1} \, 3p$~resonance.}.
This dependence of~$\sigma\X{X}(I\X{L}, \omega\X{P})$ on~$I\X{L}$ permits
a high degree of control over the absorption of FEL x~rays which
determines the probability for an atom to be core excited and thus
to enter a state of population inversion [Fig.~\ref{fig:scheme}].
The Floquet approximation is used to determine~$\sigma\X{X}(I\X{L}, \omega)$
for an atom in two-color continuous-wave~(cw) light~\cite{Buth:TX-07}.
As we employ light pulses with a duration of only a
few optical cycles, the Floquet method represents an approximation
to the non\nbh{}cw solution for finite-duration light pulses.
In Fig.~2 of Ref.~\onlinecite{Dorr:TE-95} results from the solution of the
time-dependent \hbox{Schr\"odinger} equation are compared with
Floquet theory for a three-cycle FWHM duration pulse interacting with a
hydrogen atom at a photon energy
%
%
of~$54.4 \eV$.
The parameters used to produce that figure and the situation considered
here are not identical and thus some moderate quantitative deviations
are possible.
Further details on ultrashort pulse propagation in a macroscopic medium
are examined in Ref.~\onlinecite{Gaarde:LD-11} based on
the solution of the combined Maxwell and Schr\"odinger equations.

The optical laser rapidly ionizes Rydberg electrons, \eg, in the
neon~$3p$~orbital, via multi-optical-photon absorption leading to a large
induced width for core-excited states of
approximately~$0.54 \eV$~\cite{Buth:ET-07}.
Hence, a mixture of core-excited and core-ionized atoms is found in the
gas cell.
For core-ionized neon atoms, the \xray~emission spectrum~\cite{Agren:MS-78}
peaks at~$848.66 \eV$
which is within the linewidth of the emission from core-excited
states~\cite{Oura:IS-10}.
Using a detector with a resolution that is too low to resolve the
fluorescence lines of core-excited neon, we need not
distinguish between core-excited~\cite{Oura:IS-10} and
core-ionized~\cite{Agren:MS-78} neon atoms.
The lifetime of a core-excited state is only minutely influenced by
the presence of a Rydberg electron with respect to the lifetime of a
core-ionized state~\cite{DeFanis:IE-02} and is~$2.4 \U{fs}$,
implying an XRL~linewidth of~$\Delta \omega\X{X} = 0.27
\eV$~\cite{Schmidt:ES-97}.
Furthermore, optical laser ionization, Auger decay, and
the very short coherence time of SASE FEL x~rays cause
strong decoherence, making semiclassical laser theory~\cite{Milonni:LP-10}
applicable to describe the effect of optical laser dressing on \xray~lasing.

\section{Small-signal gain}
\label{sec:SSG}

\begin{figure}
    \centering
    \includegraphics[width=\hsize,clip]{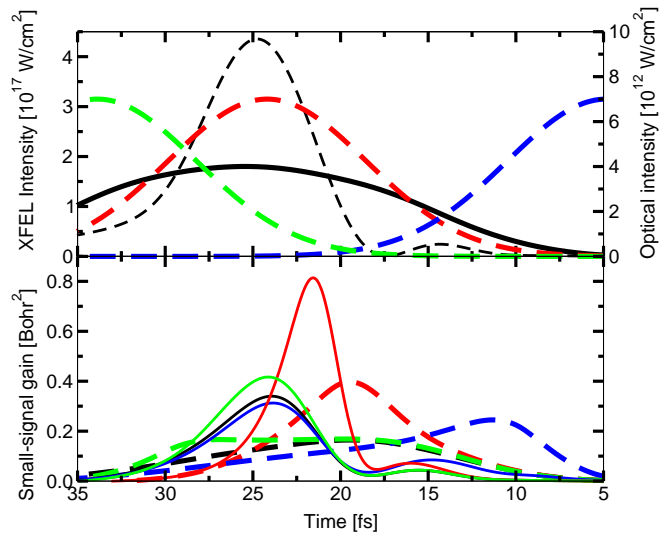}
    \caption{(Color) (a)~The dashed black line is the amplitude of a
             single SASE FEL \xray~pulse.
             The dashed red, blue, and green lines are five-optical-cycle
             %
             %
             ($13 \U{fs}$) FWHM~duration~pulses of the optical dressing laser at
             three different positions with respect to the FEL \xray~pulse.
             The solid black line is the average over 500~SASE FEL pulses.
             We assume $3 \E{13}$~x~rays in, on average, $21.5 \U{fs}$~FWHM
             long pulses, focused to a~$10 \U{\mu m}$ spot at a
             bandwidth of~$0.4 \eV$.
             (b)~The solid black line shows the single-shot SSG in
             the field-free case if the XRL is pumped with the
             dashed black SASE FEL pulse from~(a) which has a
             FWHM duration of~$7.3 \U{fs}$;
             the solid red, blue, and green lines show the single-shot
             SSG for laser dressing with the optical laser pulse of matching
             color from~(a).
             The average of the SSG over 500~SASE FEL pulses for the
             field-free case is given by the dashed black line
             and for optical laser dressing by the dashed red, blue,
             and green lines.}
    \label{fig:combined}
\end{figure}

The small-signal gain~(SSG)~\cite{Milonni:LP-10,Rohringer:AI-09,%
Rohringer:ER-10,Rohringer:LA-09,Darvasi:OC-11} represents the amplification of
spontaneously emitted x~rays on the XRL~transition in the
exponential gain regime:
\begin{equation}
  \label{eq:SSG}
  g(t) = \sigma\X{se} \, N\X{U}(t) - \sigma\X{ab} \, N\X{L}(t) \; ,
\end{equation}
where $N\X{U}(t)$~is the upper-level occupancy (or population),
$N\X{L}(t)$~is the lower-level occupancy, and $\sigma\X{se}$ and
$\sigma\X{ab}$ are the stimulated emission and absorption cross
sections, respectively, with
\begin{equation}
  \label{eq:stimabs}
  \sigma\X{se} = A\X{U \to L} \> \dfrac{2 \, \pi \, c^2}{\omega\X{X}^2 \,
    \Delta \omega\X{X}} \; , \qquad \sigma\X{ab} = \sigma\X{se} \>
    \dfrac{g\X{U}}{g\X{L}} \; ,
\end{equation}
where $A\X{U \to L} = 5.9 \E{12} \U{s^{-1}}$ is the Einstein
coefficient for spontaneous emission on the XRL~transition,
$c$~is the speed of light in vacuum, and $g\X{U} = 6$ and
$g\X{L} = 1$ are the probabilistic weights of the upper and
lower lasing levels, respectively, in our case.
The cross sections in Eq.~(\ref{eq:stimabs}) have been evaluated at the
peak of the line, assuming a Lorentzian line
shape~\cite{Rohringer:AI-09,Rohringer:ER-10,Rohringer:LA-09,Darvasi:OC-11}.
The SSG in Fig.~\ref{fig:combined}b is calculated with Eqs.~(\ref{eq:SSG})
and (\ref{eq:stimabs}) for a single atom determining the level populations
with rate equations similar to Eqs.~(\ref{eq:groundpop}),
(\ref{eq:upperlevel}), and (\ref{eq:lowerlevel}) below, however, without
the dependence on the $z$~coordinate and the influence of the XRL~light
on the level populations.
We generate SASE FEL x~rays by the partial-coherence
method~\cite{Pfeifer:PC-10,*Jiang:TC-10,*Cavaletto:RF-12} using a Gaussian
spectrum centered at~$\omega\X{P}$ with a FWHM of~$\Delta \omega\X{P}
= 0.4 \eV$.
The FEL beam is assumed to be Gaussian with a waist
of~$w_0 = 10 \U{\mu m}$~\cite{LCLS:CDR-02,Emma:FL-10}.
We place the neon atoms in a gas cell around the beam waist
with a length of~$L = 5 \U{mm}$.

In Fig.~\ref{fig:combined}a, we show optical laser pulses%
~\footnote{The beam waist of the optical laser is chosen sufficiently
large to ensure a confocal parameter~\cite{Diels:UL-06,Milonni:LP-10}
that is bigger than the length~$L$ of the gas cell.}\nocite{Diels:UL-06}
at three different positions with respect to a single FEL \xray~pulse;
in Fig.~\ref{fig:combined}b, we display the SSG for pumping
with this FEL~pulse with and without optical laser dressing
for each of the three positions.
A pronounced modulation of the SSG occurs only for the dashed red
optical laser pulse.
The SSG is modified substantially only if the optical laser pulse
overlaps with the rising flank of the FEL pulse.
Otherwise, the $K$-shell absorption cross section is modified
either too early---before the x~rays from the FEL interact with
the atoms---or too late, after a substantial fraction of the
atoms has already been destroyed, \ie, excited or ionized.

As the shape of the \xray~pulse from a SASE FEL varies on a shot-to-shot
basis~\cite{Kondratenko:GC-79,Bonifacio:CI-84,LCLS:CDR-02,Emma:FL-10},
we present SSG results averaged over 500~single-shot calculations in
Fig.~\ref{fig:combined}b for the three positions of the
optical dressing laser.
The average SASE FEL pulse over 500~shots is shown in Fig.~\ref{fig:combined}a.
The importance of the relative timing between the FEL pulse
and the optical laser pulse is also reflected in the averaged
calculations.
The average SSG increases from~$0.16 \bohr^2$ to~$0.4 \bohr^2$
for the optical laser pulse drawn as a dashed red line
in Fig.~\ref{fig:combined}a while no substantial increase
of the SSG is observed for the dashed blue and dashed green
positions of the optical laser pulse.
It is apparent in Fig.~\ref{fig:combined}b that temporal
control over the XRL is possible.
The peak of the averaged SSG is shifted towards the peak of the
optical dressing laser for each of the three cases.
The modulation of the SSG results from the increase of the
$K$-shell absorption cross section by optical laser dressing
as seen in Fig.~\ref{fig:crosssection} (bottom), and gives rise
to a modulation of the rate at which the upper-level population
and thus the population inversion are built up.
However, the optical laser does not affect processes that
destroy the population inversion, namely, the decay rates
of the upper level and the rate of valence ionization which
also reduce the occupancy of the upper level.
Therefore, optical laser dressing allows for the buildup of
a larger population inversion than in the optical-field-free case.

\section{Propagation in a medium}
\label{sec:propagation}

Based on the analysis of the SSG in Sec.~\ref{sec:SSG}, which describes
\xray~lasing in the exponential gain regime, it seems feasible to control
the XRL with a copropagating optical laser.
For a macroscopic medium, the propagation of the optical laser, the
FEL x~rays, and the XRL~x~rays need to be investigated in detail.
The ground-state population~$N\X{G}(z, t)$ at position~$z$ along the
beam axis as a function of time~$t$ is determined by the rate
equation~\cite{Darvasi:OC-11}:
\begin{equation}
  \label{eq:groundpop}
  \tfrac{\differential N\X{G}(z, t)}{\differential t} = -\dfrac{I\X{P}(z, t)}
    {\omega\X{P}} \, \bigl( \bar \sigma\X{X}(I\X{L}(z, t)) + \sigma\X{G}
    \bigr) \, N\X{G}(z, t) \; ,
\end{equation}
which depends on the FEL \xray~intensity~$I\X{P}(z, t)$ and the
weighted-average core-excitation cross section of the
optical-laser-dressed atoms,
\begin{equation}
  \label{eq:efflascross}
  \bar \sigma\X{X}(I\X{L}) = \dfrac{\Int_0^{\infty} \sigma\X{X}(I\X{L}, \omega)
    \, \tfrac{S\X{P}(z, \omega)}{\omega} \differential \omega}
    {\Int_0^{\infty} \tfrac{S\X{P}(z, \omega)}{\omega} \differential \omega}
    \; .
\end{equation}
The photoexcitation rate is the first term on the right-hand side of
Eq.~(\ref{eq:groundpop}).
It is expressed in terms of the weighted average~(\ref{eq:efflascross})
of~$\sigma\X{X}(I\X{L}, \omega)$ with the spectral
intensity~$S\X{P}(z, \omega) = \tfrac{c}{4 \pi^2} |E\X{P}(z,
\omega)|^2$ of the FEL \xray~pulse---converted to the spectral flux by dividing
by~$\omega$---that is defined with respect to the Fourier-transformed electric
field of the FEL x~rays~$E\X{P}(z, \omega)$~\cite{Diels:UL-06,Darvasi:OC-11}.
We specify the FEL photon flux by~$\tfrac{I\X{P}(z, t)}{\omega\X{P}}$
and, in doing so, approximate the FEL x~rays as monochromatic, which is
justified by~$\tfrac{\Delta \omega\X{P}}{\omega\X{P}} = 5 \E{-4} \ll 1$.
Nonetheless, the variation of the core-excitation cross
section~$\sigma\X{X}(I\X{L}, \omega)$ with respect to the
frequencies~$\omega$ in the pump pulse is accounted for
using~$\bar \sigma\X{X}(I\X{L})$.
Expression~(\ref{eq:efflascross}) accounts for the fact that different
frequency components of the FEL \xray~pulse core-excite optical-laser-dressed
atoms at varying rates due to the variation of the cross
section~$\sigma\X{X}(I\X{L}, \omega)$ around the FEL~central
frequency~$\omega\X{P}$ [Fig.~\ref{fig:crosssection}].
The second term on the right-hand side of Eq.~(\ref{eq:groundpop}) stands for
the loss of ground-state population due to valence ionization by the FEL
x~rays which is determined by the
%
%
%
cross section~$\sigma\X{G} = 23.8 \U{kilobarn}$~\cite{Cowan:TA-81,*LANL:AP-00}
where we neglect the influence of the optical laser dressing on~$\sigma\X{G}$
which is minimal because photoelectrons are ejected with a large kinetic
energy~\cite{Buth:ET-07}.

The FEL pump pulse is absorbed by the atoms in the medium in the course of the
propagation as described by Eqs.~(\ref{eq:groundpop}), (\ref{eq:ratexfel}),
(\ref{eq:upperlevel}), and (\ref{eq:lowerlevel}).
The temporal and spatial evolution of the FEL~pump
intensity~\cite{Darvasi:OC-11} is given by
\begin{eqnarray}
  \label{eq:ratexfel}
  \tfrac{\differential I\X{P}(z, t)}{\differential t} &=& -c \, n_{\#} \,
    I\X{P}(z, t) \, \bigl[ ( \bar \sigma\X{X}(I\X{L}(z, t)) +
    \sigma\X{G} ) \, N\X{G}(z, t) \\
  &&{} + \sigma\X{U} \, N\X{U}(z, t) + \sigma\X{L} \, N\X{L}(z, t) \bigr]
    - c \, \dfrac{\differential I\X{P}(z, t)}{\differential z} \; , \quad
    \nonumber
\end{eqnarray}
where $n_{\#}$~is the atomic number density of the lasing medium.
The three summands in the bracket on the right-hand side of
Eq.~(\ref{eq:ratexfel}) account for the absorption of the FEL x~rays
by atoms in the ground state and by atoms in the upper and lower levels
with the occupancies~$N\X{U}(z, t)$ and $N\X{L}(z, t)$,
where the total absorption cross sections
are~$\sigma\X{U} = 32.8 \U{kilobarn}$ for the upper and
    $\sigma\X{L} = 24.2 \U{kilobarn}$ for the lower lasing level,
respectively~\cite{Cowan:TA-81,LANL:AP-00}.
The last term on the bottom line of Eq.~(\ref{eq:ratexfel}) represents the
change of FEL pump intensity due to the traveling of the FEL pulse
in the positive $z$~direction with the speed of light in vacuum~$c$.

In order to determine the impact of optical laser dressing on the XRL,
we need to calculate its output intensity~$I\X{X}(z, t)$ for
varying FEL pulse shapes.
The occupancy of the upper lasing level~$N\X{U}(z, t)$ is found by
considering all processes that build up or remove population
in terms of the rate equation
\begin{eqnarray}
  \label{eq:upperlevel}
  \tfrac{\differential N\X{U}(z, t)}{\differential t} &=&
    \dfrac{I\X{P}(z, t)}{\omega\X{P}} \, \bar \sigma\X{X}(I\X{L}(z, t)) \,
    N\X{G}(z, t) - \dfrac{I\X{X}(z, t)}{\omega\X{X}} \, g(z, t) \nonumber \\
  &&{} - \Bigl[ \Gamma\X{A} + A\X{U \to L} + \dfrac{I\X{P}(z, t)}
    {\omega\X{P}} \, \sigma\X{U} \Bigr] \, N\X{U}(z, t) \; .
\end{eqnarray}
The first term on the right-hand side of the top line of
Eq.~(\ref{eq:upperlevel}) represents the rate at which the upper
level is populated via photoexcitation by the FEL x~rays.
The second term on the right-hand side of Eq.~(\ref{eq:upperlevel})
is the position-~$z$ and time-~$t$ dependent SSG~$g(z, t)$, defined
in analogy to Eq.~(\ref{eq:SSG}), multiplied by the flux of the
XRL~light which is treated as being
%
%
monochromatic in good approximation because~$\tfrac{\Delta \omega\X{X}}
{\omega\X{X}} =  3 \E{-4} \ll 1$.
This term specifies the rate of XRL~transitions between the lower and the
upper levels where a positive (negative) SSG leads to a reduction (increase)
of the upper-level occupancy.
The first term on the bottom line of Eq.~(\ref{eq:upperlevel}) accounts
for Auger decay of the upper lasing level with the rate~$\Gamma\X{A}
= 3.8 \E{14} \U{s^{-1}}$~\cite{Schmidt:ES-97}.
The second term describes spontaneous emission that moves population from
the upper to the lower lasing level, and the third term stands for
valence ionization by the FEL x~rays that destroy the atoms.
As the XRL~photon energy~$\omega\X{X} < \omega\X{P}$, core excitation cannot
be induced by the XRL~light.
Valence ionization by the XRL~light is omitted from
Eq.~(\ref{eq:upperlevel}), as in Refs.~\onlinecite{Rohringer:AI-09,%
Rohringer:ER-10,Rohringer:LA-09,Darvasi:OC-11}, because the valence
ionization cross section at~$\omega\X{X}$ is small and thus the
FEL x~rays will have excited or ionized the atoms before the
XRL~intensity has reached a magnitude that would make valence
ionization by the XRL~light competitive with valence
ionization by FEL x~rays.

The occupancy of the lower lasing level~$N\X{L}(z,t)$ is determined
by the rate equation
\begin{eqnarray}
  \label{eq:lowerlevel}
  \tfrac{\differential N\X{L}(z, t)}{\differential t} &=&
    \dfrac{I\X{P}(z, t)}{\omega\X{P}} \, \sigma_{\mathrm G, 2p} \, N\X{G}(z, t)
    + A\X{U \to L} \, N\X{U}(z, t) \\
   &&{} + \dfrac{I\X{X}(z, t)}{\omega\X{X}} \, g(z, t)
     - \dfrac{I\X{P}(z, t)}{\omega\X{P}} \, \sigma\X{L} \, N\X{L}(z, t) \; .
     \nonumber
\end{eqnarray}
In contrast to Eq.~(\ref{eq:upperlevel}), the population of the lower
XRL~level in the top line on the right-hand side of Eq.~(\ref{eq:lowerlevel})
is obtained, by the first term, from the valence-ionization cross
section of the $2p$~electrons of the ground-state
atoms~$\sigma_{\mathrm G, 2p} = 7.70
\U{kilobarn}$~\cite{Cowan:TA-81,LANL:AP-00} and,~%
\footnote{In specifying the first term on the right-hand side of
Eq.~(\ref{eq:lowerlevel}) involving~$\sigma_{\mathrm G, 2p}$, we
do not distinguish between valence-ionized atoms and valence-excited
atoms as lower lasing level.}
by the second term, from the transition rate from the upper to
the lower level due to spontaneous emission.
In the bottom line, the first term contains the SSG~(\ref{eq:SSG}) and has
the opposite sign as the corresponding term in Eq.~(\ref{eq:upperlevel});
the second term accounts for the destruction of the atoms in the
lower lasing level by valence ionization by the FEL
x~rays~\cite{Darvasi:OC-11}, where valence ionization by the XRL~light is again
not included.

\begin{figure}
    \centering
    \includegraphics[width=\hsize,clip]{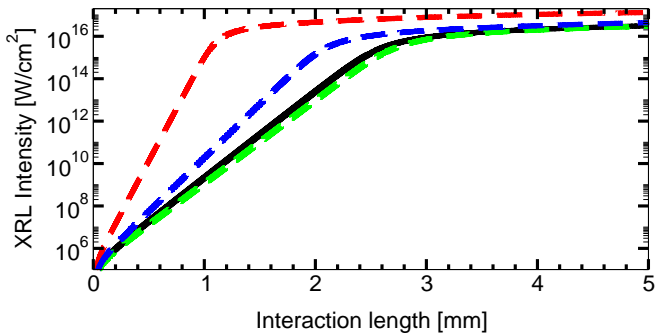}
    \caption{(Color) Dependence of the XRL~output intensity
             on the interaction length (or propagation distance)
             in a macroscopic medium.
             The XRL is pumped by the dashed black FEL pulse displayed
             in Fig.~\ref{fig:combined}a, which leads to
             the XRL~output intensity that is given by the solid black line.
             The XRL~output intensities with dressing by
             the optical laser pulses from Fig.~\ref{fig:combined}a
             are shown as dashed lines of the same color here.}
    \label{fig:output}
\end{figure}

The XRL~intensity~$I\X{X}(z, t)$ is influenced by the occupancies of the
upper and lower levels, which determine the SSG~(\ref{eq:SSG}), in
the course of the propagation in the medium.
Specifically, light with the XRL~transition frequency is attenuated
for a negative SSG and it is amplified for a positive SSG as described by
\begin{eqnarray}
  \label{eq:XRLflux}
  \tfrac{\differential I\X{X}(z, t)}{\differential t} &=& c \, n_{\#} \, \bigl[
    I\X{X}(z, t) \, g(z, t) + \omega\X{X} \, \dfrac{\Omega(z)}{4 \pi}
    \, A\X{U \to L} \, N\X{U}(z, t) \bigr] \nonumber \\
  &&{} - c \, \dfrac{\differential I\X{X}(z, t)}{\differential z} \; .
\end{eqnarray}
The first term on the right-hand side of Eq.~(\ref{eq:XRLflux})
describes stimulated emission and absorption of light from the XRL~pulse
via the SSG [Eq.~(\ref{eq:SSG})], whereas the second term represents
the rate of spontaneous emission which initiates \xray~lasing in the
forward direction because the XRL~intensity is zero in the beginning.
Only photons emitted into the solid angle
\begin{equation}
  \Omega(z) = 2 \, \pi \, \Bigl( 1 - \dfrac{L - z}{\sqrt{w_0^2 + (L - z)^2}}
    \Bigr)
\end{equation}
contribute to \xray~lasing because only these photons stay completely within
the XRL~medium of length~$L$ and beam waist~$w_0$ for the remainder of the
propagation~\cite{Rohringer:AI-09,Rohringer:ER-10,Rohringer:LA-09,%
Darvasi:OC-11}.
The bottom line of Eq.~(\ref{eq:XRLflux}) stands for the propagation of
the XRL~light traveling in the positive $z$~direction.

We use a fourth-order Runge-Kutta algorithm~\cite{Milonni:LP-10}
to solve the linear system of ordinary first-order differential
equations constituted by Eqs.~(\ref{eq:groundpop}), (\ref{eq:upperlevel}),
and (\ref{eq:lowerlevel}), transformed to a reference frame traveling
with the speed of light.
This gives the temporal evolution of the occupancy of the ground
state~$N\X{G}(z, t)$ and the upper~$N\X{U}(z, t)$ and the lower~$N\X{L}(z, t)$
lasing levels for position~$z$ at time~$t$.
Equations~(\ref{eq:ratexfel}) and (\ref{eq:XRLflux}) are solved through
first-order forward stepping to find the pump~$I\X{P}(z, t)$ and
XRL~$I\X{X}(z,t)$ intensities~\cite{Milonni:LP-10,Darvasi:OC-11}.

We apply Eqs.~(\ref{eq:groundpop}), (\ref{eq:ratexfel}), (\ref{eq:upperlevel}),
(\ref{eq:lowerlevel}), and (\ref{eq:XRLflux}) to our
XRL~scheme and calculate~$I\X{X}(z,t)$ depending on the interaction length
(or propagation distance), which is the distance that the FEL, optical
laser, and XRL~pulses have propagated in a gas cell with~$n_\# = 10^{19}
\U{cm^{-3}}$.
\emph{Nota bene}, the $z$~coordinate in~$I\X{X}(z,t)$ does not represent the
propagation distance but describes the longitudinal spatial profile of the
XRL~pulses for an instant in time~$t$.
In Fig.~\ref{fig:output}, we display~$I\X{X}(z,t)$ of an XRL pumped by
the FEL x~rays only and with additional optical laser dressing for
the three cases shown in Fig.~\ref{fig:combined}a.
These are single-shot calculations of the XRL~output intensity
for the single-shot SSGs displayed in Fig.~\ref{fig:combined}b.
Significant modulation of the XRL~intensity is only provided by the
dashed red, optical laser pulses displayed in Fig.~\ref{fig:combined}a.
The other optical laser pulses do not overlap with the rising flank
of the FEL pump pulse and thus only slightly alter the XRL intensity
with interaction length.

\begin{figure}
    \centering
    \includegraphics[width=\hsize,clip]{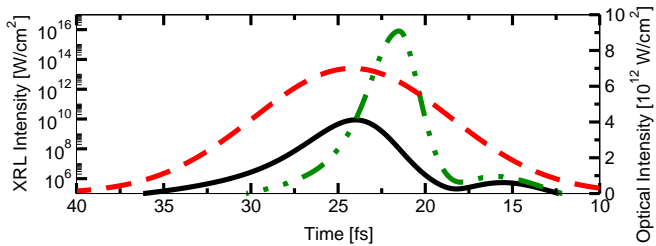}
    \caption{(Color online) The temporal profile of the XRL pulses
             after saturation.
             The XRL is pumped with the dashed black FEL pulse from
             Fig.~\ref{fig:combined}a.
             The solid black line corresponds to the FEL-\xray-only
             case and the dot-dashed green line displays the
             optical laser-dressed case by the dashed red optical
             laser pulse from Fig.~\ref{fig:combined}a, which is shown
             here as well.}
    \label{fig:pulses}
\end{figure}

The temporal profiles of the XRL pulse after propagation through the medium
for the FEL-\xray-only case and for optical laser dressing with the
dashed red optical laser pulse are shown in Fig.~\ref{fig:pulses}.
The small peak at the front of the XRL pulses is ascribed
to emission processes from a small population inversion
caused by the small peak in the FEL pulse found at
approximately~$15 \U{fs}$ in Fig.~\ref{fig:combined}a.
The comparison of the solid black and dot-dashed green lines reveals
that optical laser dressing leads to an increase by about
six orders of magnitude of the XRL's peak intensity.
Hence, the XRL's output intensity can be controlled via
optical laser dressing.
The FWHM~duration of the XRL pulse from the FEL-\xray-only
case in Fig.~\ref{fig:pulses} is compressed from~$2.0 \U{fs}$ down
to~$0.7 \U{fs}$ if the dashed red optical laser pulse is used due to
a faster buildup of population inversion in the latter case compared with
the former case.
Furthermore, Fig.~\ref{fig:pulses} reveals that the XRL~pulse is
synchronized to the optical laser pulse with a time jitter of better
than~$5 \U{fs}$.
At peak FEL intensities reaching~$4 \E{17} \U{W/cm^2}$, the first
few femtoseconds of the FEL pulse induce a complete population inversion,
\ie, all neon atoms are either core excited afterwards or valence ionized,
such that the remainder of the pulse passes through the gas without
further interaction with core electrons.
The energy efficiency of the XRL, \ie, the number of FEL photons that
actually contribute to core excitation of the macroscopic medium, can
be improved by reducing the FEL's peak intensity or shortening its pulses,
however, reducing the FEL peak intensity leads to longer and less
intense XRL~pulses~\cite{Darvasi:OC-11}.

\section{Conclusion}
\label{sec:conclusion}

We propose an inner-shell XRL~scheme that
allows one to produce controlled, fully coherent, single-peak
\xray~pulses with intensities above~$10^{16} \U{W/cm^2}$ which
are synchronized to an optical laser pulse with femtosecond accuracy.
In our scheme, optical-laser-controlled core excitation produces
a state of population inversion;
spontaneously emitted x~rays which propagate along the beam axis
of the FEL pump x~rays are then amplified by stimulated emission
leading to \xray~lasing.

Our XRL scheme can be modified in such a way that it works inversely to
what has been discussed in this work by shifting the central
frequency~$\omega\X{P}$ of the FEL x~rays to the $1s \to 1s^{-1} \,
3p$~resonance.
Then, the optical laser dressing suppresses the $K$-shell
absorption cross section by core electrons of neon as seen in
Fig.~\ref{fig:crosssection} and x~rays from the FEL are only
absorbed efficiently when optical light is not present.
This can be used to produce a short XRL~pulse from a long
FEL \xray~pulse by applying a long optical laser pulse with
an interruption of a few femtoseconds.

Our proposed XRL goes beyond what can be realized at present-day \xray~FELs
and will stimulate feasible and attractive future \xray~science.
Especially challenging for an experimental realization of our scheme
is the achievement of~$3 \E{13}$~FEL photons in a
$0.4 \eV$~bandwidth interval which is not currently possible,
for example, at the soft \xray~instrument of LCLS for which
estimates reveal that about~$2 \E{10}$~photons are available
under such conditions~\cite{Schlotter:SX-12}.
Our idea thus opens up perspectives for future two-color pump-probe
experiments.

The optically controlled XRL offers a different way to measure the time jitter
between FEL~radiation and an optical laser that complements existing
methods at present-day facilities~\cite{Bionta:SE-11,*Schorb:XR-12,%
*Harmand:FF-13}.
Namely, choosing a length of the gas cell of~$1.1 \U{mm}$
[Fig.~\ref{fig:output}] and placing a suitable
filter~\cite{Henke:XR-93,*Henke:XI-10} behind
the gas cell that absorbs an intensity of up to~$10^{11} \U{W/cm^2}$,
the output of the FEL-\xray-only XRL can be entirely absorbed.
Then, XRL output is only generated when the optical laser pulse
overlaps with the front of the FEL pulse.
This allows one to determine the overlap between x~rays
and the optical laser with femtosecond precision.
Also the FEL x~rays are damped down by many orders of magnitude by
the propagation in the macroscopic medium and by the filter afterwards.
Since the photon energies of the XRL and the FEL are comparable, the filter
attenuates both beams by similar amounts.
In order to distinguish between XRL light and FEL x~rays, either a
spectrometer is required to resolve the different photon energies or
an inclined-beam experimental setup needs to ensure a spatial separation
of both beams after propagation through the medium.

Multicolor \xray~lasing on several transitions, as predicted for a
core-ionization-pumped XRL~\cite{Rohringer:AI-09,Rohringer:ER-10,%
Rohringer:LA-09}, is substantially suppressed by using core excitations
because the chosen FEL photon energy of~$\omega\X{P} = 868.2 \eV$ is
far off resonant, \eg, with the $1s \to 1s^{-1} \, 2p$ and $1s \to
1s^{-1} \, 3p$~resonances in singly-charged neon cations at transition
energies of~$848.66 \eV$ and $885.8 \eV$, respectively~\cite{Oura:IS-10}.
If one is only interested in suppressing multicolor
\xray~lasing, then the FEL \xray~energy may be tuned to the
$1s \to 1s^{-1} \, 3p$~resonance.
Such a core-excitation-pumped XRL has the added benefit of
a larger SSG [Eq.~(\ref{eq:SSG})] compared with a core-ionization-pumped XRL.
This gives rise to a more rapid increase of the XRL output intensity with
propagation distance in the macroscopic medium [Fig.~\ref{fig:output}],
leading to a higher achievable saturation intensity of the~XRL.

So far we have considered a constant transverse intensity profile of the
optical laser and the FEL x~rays.
By choosing a transverse intensity profile for the optical laser pulse which is
not constant~\cite{Milonni:LP-10} allows one to imprint such a profile
onto the transverse intensity profile of the XRL~pulse, \ie, transverse
pulse shaping is facilitated.

\begin{acknowledgments}
We would like to thank Stefano M.~Cavaletto for fruitful discussions.
C.B.~was supported by the Chemical Sciences, Geosciences, and
Biosciences Division of the Office of Basic Energy Sciences,
Office of Science, U.S.~Department of Energy, under
Contract No.~DE-AC02-06CH11357.
\end{acknowledgments}

\end{document}